# *Ferromagnet/Insulator/Ferroelectric nanometer multilayer for multiferroictronics*

*Hui Zhang*[*]

*School of Materials Science and Engineering, South China University of Technology, Guangzhou 510640, People's Republic of China*

Abstract

We have investigated the field-induced-changes in both the magnetization and the polarization in ferromagnet/Insulator/ferroelectric (FM/I/FE) multilayer by following both the Stoner-Wohlfarth (SW) model and the Landau theory. It has been found that with the stresses introduced in the FM/I/FE structure by the fields, both the magnetization and the polarization states can be significantly modified and the combination of their states can be of multiple states. These results demonstrate the feasibility of combining both the spintronics and the ferroelectrics into the multiferroictronics.

## I. INTRODUCTION

In recent years, more attention has been paid to multiferroic materials which consist of ferroelectricity, ferroelasticity and ferromagnetism [1~3]. But this also has brought our interest to a question of whether there is a possibility of combining both the spintronics and the ferroelectrics into multiferroic electronics (multiferroictronics). For example, for a simple structure like ferromagnet/insulating layer/ferroelectric (FM/I/FE) multilayer, FM layer of this structure can carry the spin-polarized current which can be used for spin-dependent devices when a soft metallic ferromagnet is incorporated as a FM layer [4~6]. On the other hand, FE layer can be used for ferroelectric devices with its dielectric property and response to an applied voltage [7~10]. For this structure, if FM layer is magnetostrictive, there is also another advantage that the magnetization of FM layer can indirectly correlate with the polarization of FE layer through the stresses, and it may show different combinations of both the magnetization and the polarization states which can be used for multi-bit memory technology. The indirect coupling between the polarization of FE layer and the magnetization of FM layer can be produced with the stresses introduced by a magnetic field or an electric field in the structure. Such couplings have been experimentally confirmed in $CoFe_2O_4/BaTiO_3$ [11], $La_{0.67}Sr_{0.33}MnO_3/BaTiO_3$ [12] and $CoFe_2O_4/Pb(Mg_{1/3}Nb_{2/3})_{0.7}Ti_{0.3}O_3$ [13] epitaxial films or heterostructures. Our previous numerical investigation has also shown that with the stresses introduced in $Tb_{0.3}Dy_{0.7}Fe_2$/PZT laminate composite by a magnetic field or an electric field both the magnetization and the polarization states can be significantly modified [14]. Therefore, it is expected that more interesting results can also be obtained for the FM/I/FE nanometer multilayer.

[*]Email: zhope@scut.edu.cn

In this paper, based on both the Stoner-Wohlfarth (SW) model and the Landau theory, we have investigated the FM/I/FE structure where both soft ferromagnetic metal with a moderate saturation magnetostriction and ferroelectric layer are used. The numerical results have shown that with the strains introduced by the magnetic field or electric field in the structure, both the magnetization of FM layer and the polarization of FE layer can be significantly changed.

## II. THEORETICAL MODELING

Fig. 1. (a) Schematic for ferromagnet/insulating layer/ferroelectric (FM/I/FE) multilayer. (b), (c) The electrostrictive and electromechanical behaviors for FE layer under the applied electric field $E$ along the $z$ direction or the magnetic field induced tensile stress along the $y$ direction $\sigma_H$, respectively. (d) The magnetostrictive and magnetomechanical behaviors for FM layer with the magnetic field $H$ in the $y$ direction and the uniaxial compressive stress $\sigma_E$ along the $x$ direction present. (e) The geometric relationship between the saturation magnetization $M_s$, magnetocrystalline

anisotropy field $H_K$, and stress-induced anisotropy field $H_\sigma$ for FM layer with the saturation magnetostriction $\lambda_s > 0$. The solid lines and dashed lines show the responses of both FM layer and FE layer without and with applied fields, respectively.

Figure 1(a) shows a schematic for ferromagnet/insulating layer/ ferroelectric (FM/I/FE) multilayer. FE layer can be strained by an applied electric field in the $z$ direction or an external in-plane stress (as shown Figs. 1(b) and (c)). For a 2D clamping FE layer shown in Fig.1 (a), the strains caused by both insulating layer and FM layer can induce the phase transition of FE layer [15~16]. In our case, FE layer is initially set to be in the $c$ phase ($P_x$=$P_y$=0, $P_z$≠0). Under the tensile strain, this $c$ phase becomes unstable and the phase transition to the $aa$ phase ($P_x$≠$P_y$≠0, $P_z$=0) occurs. With the applied electric field present, for FE layer $P_z$ increases, leading to the elongation in the $z$ direction and the contraction in both the $x$ and $y$ directions.

For FM layer, the magnetization is assumed to lie in the plane, and the magnetocrystalline anisotropy field is set to be along the $x$ direction (as shown Figs. 1(d) and (e)). As a magnetic field is applied along the $y$ direction, FM layer can be elongated in the $y$ direction. Under the action of a stress along the $x$ direction, the coupling of the magnetization with the stress can create the stress-induced anisotropy field $H_\sigma$ along the $x$ direction or the $y$ direction, which is determined by the sign of the product of $\lambda_s\sigma_E$. Thus the direction of magnetization can be altered.

For FM layer with the uniaxial anisotropy, the free energy per unit volume can be written as:

$$F_M = -\mu_0 H_y M_s \cos(\frac{\pi}{2} - \theta) + K_1 \sin^2(\theta) + \frac{3}{2}\lambda_s \sigma_E \sin^2(\theta) \quad (1)$$

where the first term is the field energy, the second term is the magnetocrystalline anisotropy energy, and the third term is the stress-induced anisotropy energy. Note that the expression of the stress-induced anisotropy energy for isotropic materials is used for simplicity. In Eq. (1), $\theta$ is the angle between the magnetization and the magnetocrystalline anisotropy field, and $K_1$ is the magnetocrystalline anisotropy constant. For single domains, the magnetostriction $\lambda_\theta$ can be given by $\lambda_\theta$=3/2$\lambda_s$ [$\cos^2(\pi/2-\theta)$-1/3]. Assuming that FM layer consists of non-interacting single domains, the magnetostrictive behaviors for the two types of domains existing in FM layer are the same. Therefore the field induced change in magnetostriction $\Delta\lambda_M$ for FM layer can be written as:

$$\Delta\lambda_M = \lambda_\theta - \lambda_i \tag{2}$$

where $\lambda_\theta$ and $\lambda_i$ are defined as the magnetostrictions corresponding to the final and initial magnetization states, respectively.

The components of magnetization for FM layer along the $y$ direction $M_y/M_s$ can be expressed as:

$$M_y / M_s = \sin(\theta) \tag{3}$$

For single domain FE layer with the perovskite crystal structure, the free energy per unit volume can be written as [16]:

$$F_P = \left( a_1 - \Delta\lambda_H \frac{2Q_{12}}{s_{11}+s_{12}} \right) P^2 + \left( a_{11} + \frac{Q_{12}^2}{s_{11}+s_{12}} \right) P^4 + a_{111}P^6 + \frac{\Delta\lambda_H{}^2}{s_{11}+s_{12}} - EP \tag{4}$$

where $a_1$, $a_{11}$ , and $a_{111}$ are the dielectric stiffness coefficients at constant strain, $Q_{12}$ is the electrostrictive coupling coefficient, $s_{11}$ and $s_{12}$ are the elastic compliances at constant polarization, $P$ is the polarization, $\Delta\lambda_H$ is the in-plane strain induced by the magnetic field, and $E$ is the electric field. Neglecting the contribution from the piezoelectric effect, the field induced changes in electrostriction $\Delta\lambda_P$ can be given by [7]:

$$\Delta\lambda_E = Q_{12}\left(P^2 - P_i^2\right) \tag{5}$$

where $P$ and $P_i$ are the electric polarization values corresponding to the final and initial polarization states, respectively.

Assuming that the axial tensile strain $\Delta\lambda$ along the $y$ direction is introduced in FM layer as a magnetic field is applied along the $y$ direction, the strain in FE layer is also $\Delta\lambda$. With this structure strained, FE layer is subjected to the tensile stress from FM layer, and FM layer subjected to the compressive stress from FE layer. This is also the case for the electric field induced strain. If $\mu_M$ and $\mu_P$ represent the Young's modulus of FM layer and FE layer, respectively, and $t_M$ and $t_P$ the thicknesses, respectively, the axial stresses caused by the fields can be expressed as [17]:

$$\begin{aligned} \sigma_E &= \frac{t_M \mu_M \mu_P}{\left(t_P \mu_P + t_M \mu_M\right)}\left(\Delta\lambda_M - \Delta\lambda_P^{\sigma}\right) \\ \sigma_H &= \frac{t_P \mu_P \mu_M}{\left(t_P \mu_P + t_M \mu_M\right)}\left(\Delta\lambda_P - \Delta\lambda_M^{\sigma}\right) \end{aligned} \tag{6}$$

where $\Delta\lambda_M$ and $\Delta\lambda_P$ are the strains induced by the magnetic field and the electric field, respectively, and $\Delta\lambda_M^{\sigma}$ and $\Delta\lambda_P^{\sigma}$are the strains due to the magnetomechanical effect and the electromechanical effect, respectively. For the multilayer, the strains caused by the fields can be given by:

$$\begin{aligned} \Delta\lambda_E &= \frac{\sigma_H}{\mu_M} + \Delta\lambda_M^{\sigma} = \frac{t_P \mu_P \Delta\lambda_P + t_M \mu_M \Delta\lambda_M^{\sigma}}{\left(t_P \mu_P + t_M \mu_M\right)} \\ \Delta\lambda_H &= \frac{\sigma_E}{\mu_P} + \Delta\lambda_P^{\sigma} = \frac{t_M \mu_M \Delta\lambda_M + t_P \mu_P \Delta\lambda_P^{\sigma}}{\left(t_P \mu_P + t_M \mu_M\right)} \end{aligned} \tag{7}$$

In Eqs. (6) and (7), $\Delta\lambda_M$ and $\Delta\lambda_P^{\sigma}$ or $\Delta\lambda_P$ and $\Delta\lambda_M^{\sigma}$ are mutually dependent. Therefore it is very difficult to obtain the appropriate boundary conditions for the calculations of their values. Here the values of $\Delta\lambda_P^{\sigma}$ and $\Delta\lambda_M^{\sigma}$ are approximately replaced with those for $\Delta\lambda_M$ and $\Delta\lambda_P$, respectively.

It has been seen from Eqs. (2) and (5) that these strains are strongly dependent on ferromagnetic states or ferroelectric states. For both FM and FE layers, both the magnetization and polarization states can be calculated from Eqs. (1) and (4) by following the conventional free energy minimization procedure, respectively. After knowing these states, the field-induced strains can be obtained from Eqs. (2) and (5). For simplicity, the field-induced strains without the restraint from other layers are used. Note that in real structures these strains can be induced only by the very strong fields. For both FM and FE layers with the uniaxial anisotropy, there exist two types of domains which have the same initial and final states, and the same field-induced strains. Therefore, the fractional volume of 1 is used in Eqs. (2) and (5). In the computation, for FM layer, $\mu_0 M_s$ = 1.5 Tesla, $K_1$ =4×10$^3$ J/m$^3$, $\lambda_s$=300×10$^{-6}$, and $\mu_M$=10×10$^{10}$ N/m$^2$, with reference to experimental data of CoFe alloys [18,19]. For FE layer, $a_1$= 3.3(*100*-110)×10$^5$ C$^{-2}$.m$^2$.N, $a_{11}$=3.6(*100*-175)×10$^6$ C$^{-4}$.m$^6$.N, $a_{111}$=6.6×10$^9$ C$^{-6}$.m$^{10}$.N, $Q_{12}$=-0.043 C$^{-2}$.m$^4$, $s_{11}$=8.3×10$^{-12}$ m$^2$/N, $s_{12}$=-2.7×10$^{-12}$ m$^2$/N, and $\mu_P$= 10×10$^{10}$ N/m$^2$, with reference to experimental data of $BaTiO_3$ [15]. For FM layer and FE layer, $t_M$=$t_P$=30 nm.

## III. RESULTS AND DISCUSSION

Fig. 2 (color online). Dependence of the components of polarization of FE layer $P_z$ on the tensile strain of FM layer caused by the magnetic field along the *y* direction $\Delta\lambda_H$. The inset shows the dependence of the tensile strain of FM layer $\Delta\lambda_H$ on the magnetic field $H_y$. Note that the temperature is 100 °C for a small value of $|a_1|$.

For FM/I/FE multilayer, the magnetostrictive strains caused by the magnetic field in FM layer can make FE layer strained. For FE layer, this strain can couple with the polarization, and then lead to a significant change in polarization. Figure 2 shows the dependence of the components of polarization of FE layer $P_z$ on the tensile strain of FM layer caused by the magnetic field along the *y* direction $\Delta\lambda_H$. As the magnetic field increases, the magnetization of FM layer tends to be rotated toward the *y* direction. Then the structure is elongated along the *y* direction, leading to a tensile strain in the *y* direction $\Delta\lambda_H$ (shown in the inset in Fig. 2). This tensile strain is also introduced in FE layer. With the tensile strain $\Delta\lambda_H$ increasing, the components of polarization along the $z$ direction for FE layer $P_z$ can decrease to 0. The reason lies in the phase transition in FE layer caused by the strain. For FE layer of $BaTiO_3$, the tensile strain can make the $c$ phase ($P_x$=$P_y$=0, $P_z$≠0) unstable, and the phase transition to the $aa$ phase ($P_x$=$P_y$≠0, $P_z$=0) occurs. This means that the strains caused by the magnetic field can make FE layer in the polarization states corresponding to high (~±$P_i$) and low (~0) values. Under the action of bias electric field, the polarization states of FE layer can be further modified.

Fig. 3 (color online). Dependence of the components of magnetization of FM layer $M_y/M_s$ on the compressive in-plane strain of FE layer caused by the electric field $\Delta\lambda_E$. The inset shows the dependence of the compressive in-plane strain of FE layer $\Delta\lambda_E$ on the electric field $E$ along the $z$ direction.

Conversely, the magnetization of FM layer can be significantly modified by the electric field. Figure 3 shows the dependence of the components of magnetization of FM layer $M_y/M_s$ on the compressive in-plane strain of FE layer caused by the electric field $\Delta\lambda_E$. It can be seen from Fig. 3 that FE layer can be elongated along the $z$ direction with the electric field increasing and contracted in the $x$-$y$ plane. The strain in FE layer is compressive and can be calculated according to Eq. (5). This compressive strain (the compressive stress) also is introduced in FM layer. With the stress absent, the direction of magnetization in FM layer is determined by the magnetocrystalline anisotropy which in our case creates easy directions parallel and antiparallel to the +$x$ direction. In the presence of the strain, the strain can couple with the magnetization, leading to the stress-induced anisotropy in FM layer [20~22]. Thus, the direction of magnetization in FM layer will be determined by the combination of both the magnetocrystalline and stress-induced anisotropies. When the stress is large enough to overcome the influence exerted by the magnetocrystalline anisotropy, the magnetization in FM layer is perpendicular to the stress direction (the $x$ direction). Then the values for $M_y/M_s$ can change to 0 from ±1. With the bias magnetic field present, the magnetization is rotated toward the field direction.

More interestingly, for such a simple FM/I/FE structure, the combination of both the magnetization and polarization states under the magnetic field or the electric field can be of multiple states (shown in Table I). This means that the structure can pose the potential applications in the future multi-bit memory technology.

Table I. The combination of both the magnetization and polarization states under the magnetic field $H_y$ or the electric field $E$.

| *No.* | $M_y$ | $P$ | $\pm H_y$ | $\pm E$ |
|---|---|---|---|---|
| 1 | 0 | $\pm P_i$ | 0 | 0 |
| 2 | $\pm M_s$ | 0 | 1 | 0 |
| 3 | $\pm M_s$ | $\pm P_z$ | 0 | 1 |

The above results have shown that both the magnetization of FM layer and the polarization of FE layer can be modified with the strains introduced by the combination of both the magnetic field and the electric field. These state changes in FM/I/FE multilayer can be known by detecting the changes in magnetoresistance [6] and electroresistance (polarization) [23]. For our FM/I/FE structure, the strains caused by the magnetic field or the electric field on the order of $300{\sim}500\times10^{-6}$ are large enough to induce appropriate changes in both the magnetization and the polarization. The choice for the soft metallic FM requires both small values of magnetocrystalline anisotropy and appropriate values of magnetostriction ($\sim300\times10^{-6}$). For real applications, large magnetoresistive ratios are also required. The alloys with large anisotropic magnetoresistance ratio such as CoFe alloy may be the potential candidates [6,18,19]. In addition, it is convenient to replace FM layer with giant magnetoresistive spin-valves (GMR-SVs) or tunneling magnetoresistive junction (TMR) structures with moderate magnetostriction [4,5].

For FE layer, appropriate ferroelectric materials must be chosen for large magnetic field induced changes in polarization because of small values of $\Delta\lambda_H$. As shown in Eq. (4), the values for $a_1^*=a_1-2\Delta\lambda_H\times Q_{12}/(s_{11}+s_{12})$ can become positive from the originally negative value with the increase of in-plane strain, and then the phase transition of FE layer occurs. For a small value of $|a_1|$, the magnetic field induced strains of ~$150\times10^{-6}$ may induce a detectable change in polarization (as shown in Fig. 2). For $BaTiO_3$ ferroelectric, we have used the near Curie value of $a_1=3.3(100\text{-}110)\times10^5$ to ensure the transition of the *c* phase to the *aa* phase. For large values of $|a_1|$, the magnetic field induced strains are comparatively small so that this phase transition does not occur and there is no significant change in polarization.

Our numerical results are in agreement with the recent experimental observations in Ni/$BaTiO_3$ [24], Fe/$BaTiO_3$ [25], and $Co_{90}Fe_{10}$/$BiFeO_3$ [26] heterostructures. For example, the magnetization reversal caused by the electric field has been observed in $Co_{90}Fe_{10}$/$BiFeO_3$ heterostructures. For these soft metallic alloys, their saturation magnetostrictions (30~$150\times10^{-6}$) are very small and the magnetization can be modified only by the large strains of FE layer. But for other magnetostrictive materials with large saturation magnetostriction such as Terfenol-D [27,28], FeGa [29,30] and $CoFeO_3$ [11,13], the experimental data have shown strong evidence of magnetic field induced changes in polarization. However, a lot of work must be done for both the appropriate soft magnetostrictive metallic alloys and ferroelectric materials for future applications.

## III. CONCLUSIONS

Our results have shown that for FM/I/FE multilayer the indirect coupling between the magnetization of FM layer and the polarization of FE layer can be realized through the strains introduced by the magnetic field or the electric field in the structure. The combination of both the magnetization and polarization states can be of multiple states.

## IV. ACKNOWLEDGEMENTS

This work is supported by the Fundamental Research Funds for the Central Universities.

Fig. 1. (a) Schematic for ferromagnet/insulating layer/ferroelectric (FM/I/FE) multilayer. (b), (c) The electrostrictive and electromechanical behaviors for FE layer under the applied electric field $E$ along the $z$ direction or the magnetic field induced tensile stress along the $y$ direction $\sigma_H$, respectively. (d) The magnetostrictive and magnetomechanical behaviors for FM layer with the magnetic field $H$ in the $y$ direction and the uniaxial compressive stress $\sigma_E$ along the $x$ direction present. (e) The geometric relationship between the saturation magnetization $M_s$, magnetocrystalline anisotropy field $H_K$, and stress-induced anisotropy field $H_\sigma$ for FM layer with the saturation magnetostriction $\lambda_s$ >0. The solid lines and dashed lines show the responses of both FM layer and FE layer without and with applied fields, respectively.

Fig. 2 (color online). Dependence of the components of polarization of FE layer $P_z$ on the tensile strain of FM layer caused by the magnetic field along the $y$ direction $\Delta\lambda_H$. The inset shows the dependence of the tensile strain of FM layer $\Delta\lambda_H$ on the magnetic field $H_y$. Note that the temperature is 100 °C for a small value of $|a_1|$.

Fig. 3 (color online). Dependence of the components of magnetization of FM layer $M_y/M_s$ on the compressive in-plane strain of FE layer caused by the electric field $\Delta\lambda_E$. The inset shows the dependence of the compressive in-plane strain of FE layer $\Delta\lambda_E$ on the electric field $E$ along the $z$ direction.

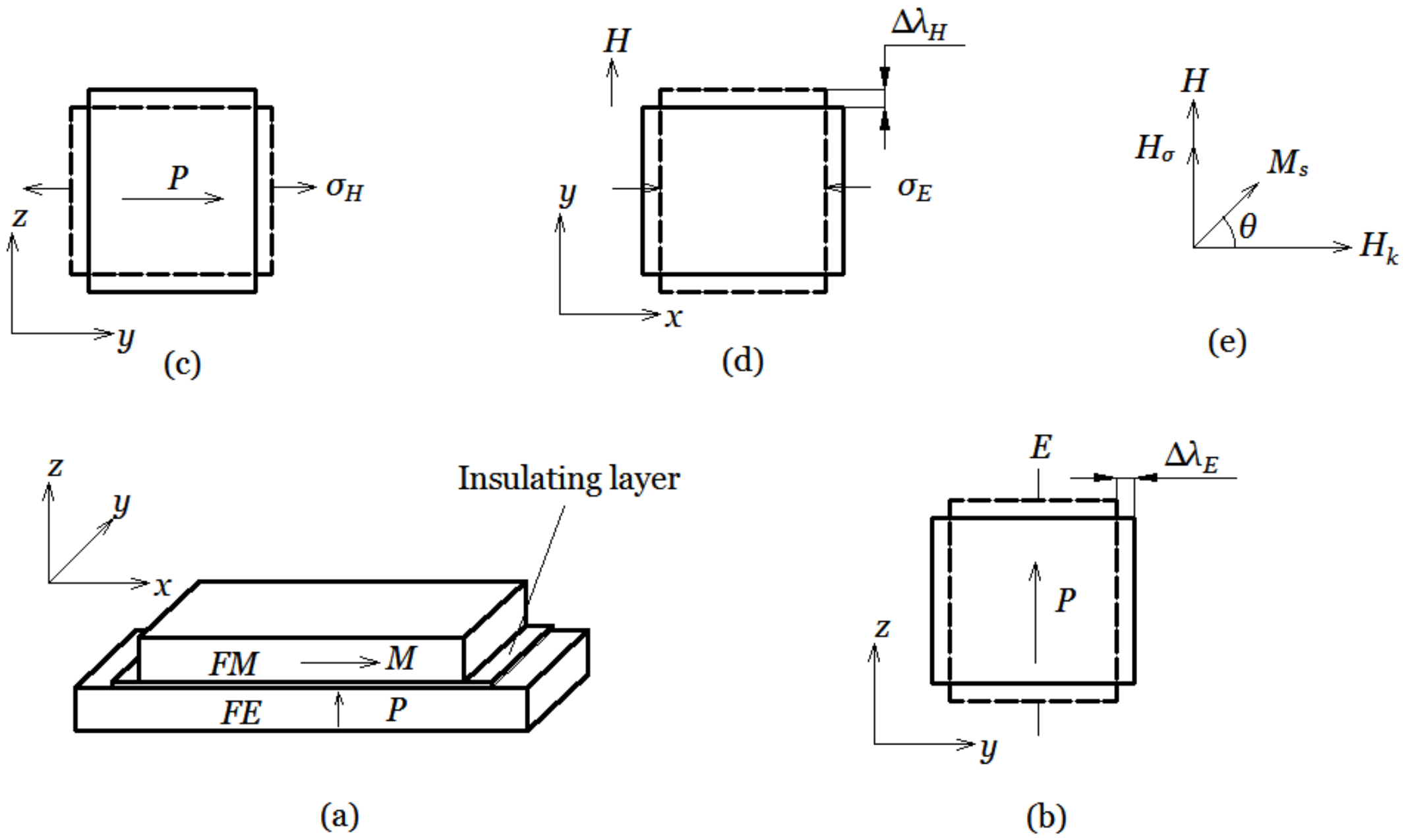

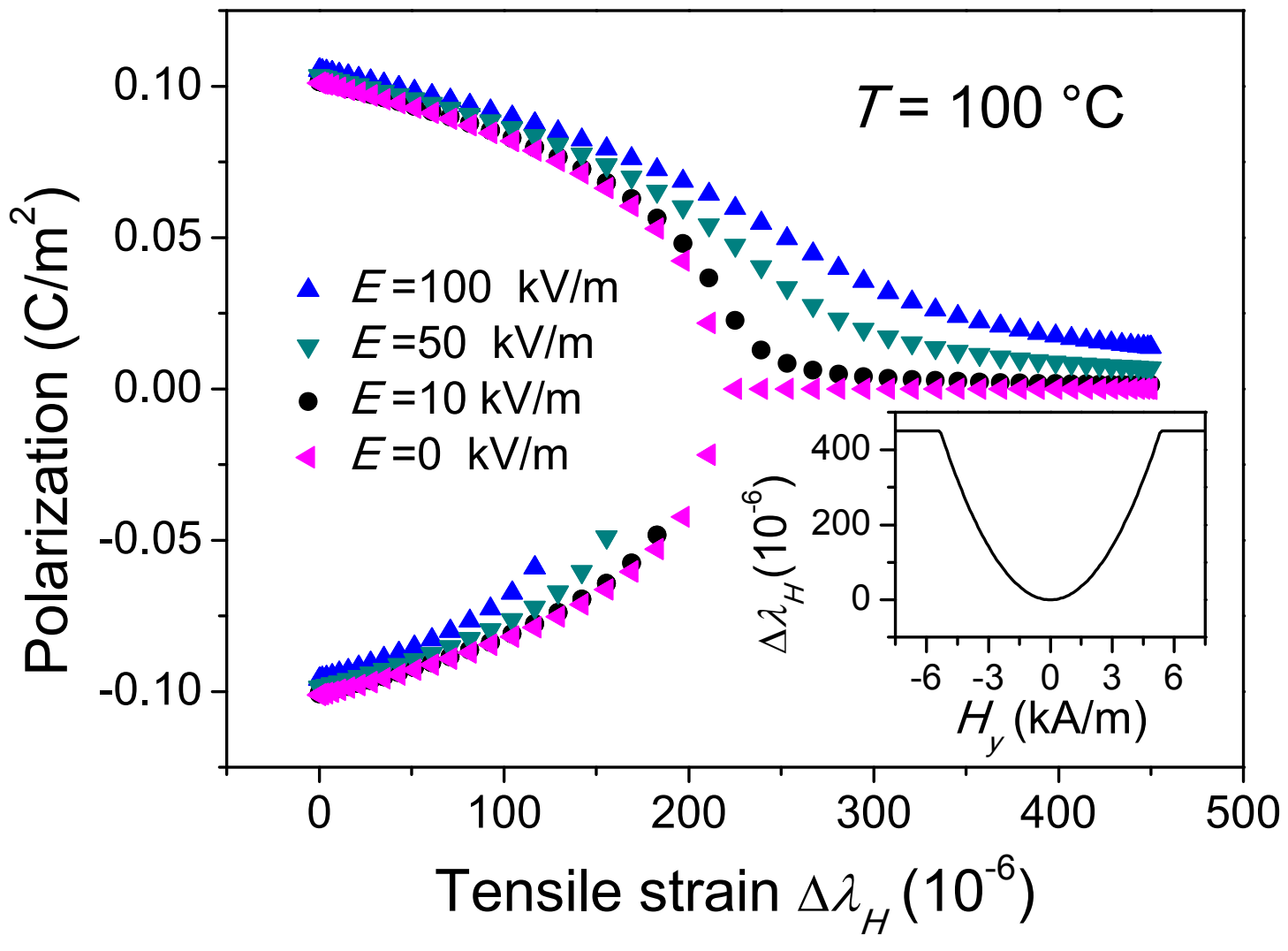

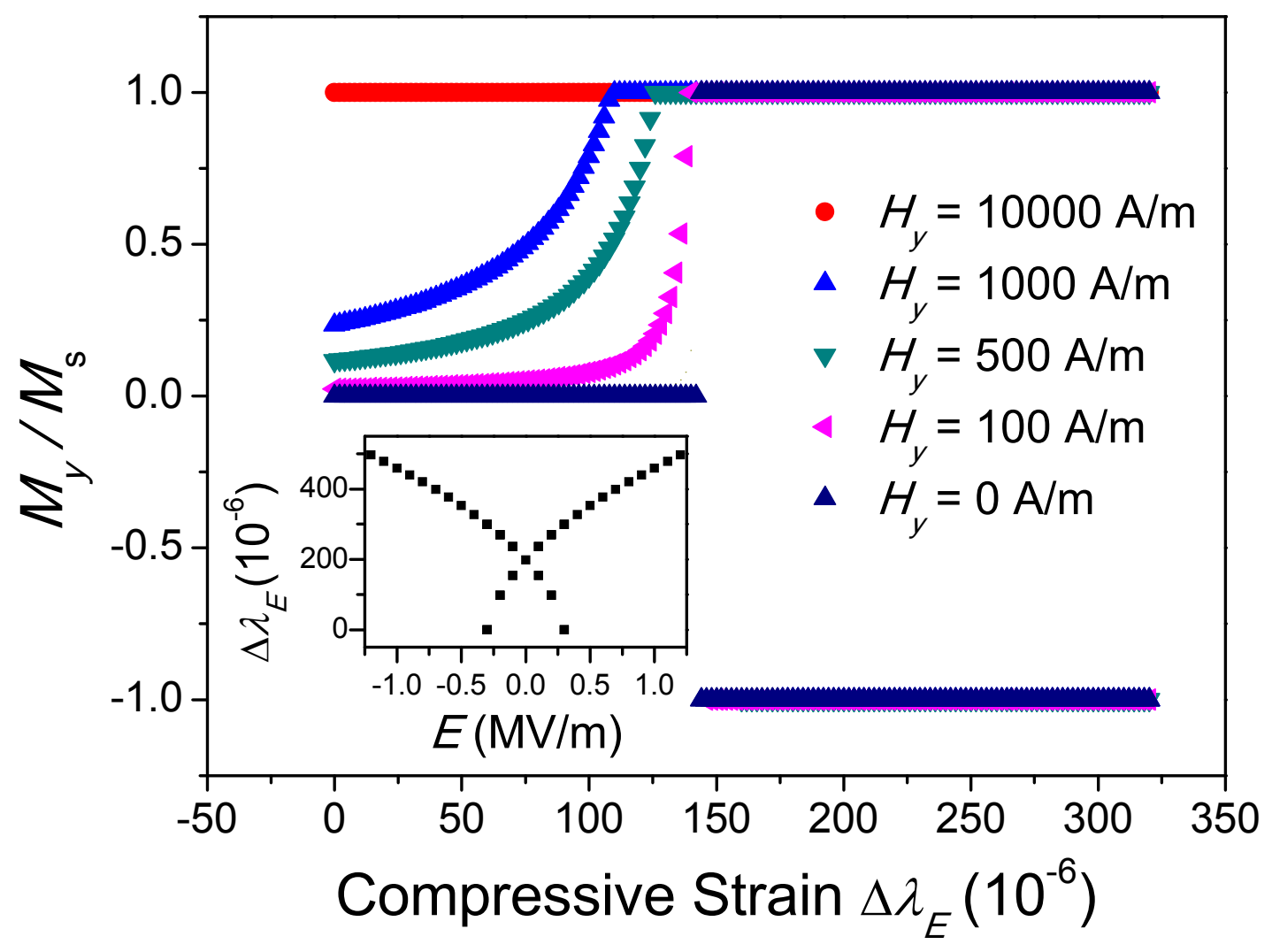